\documentclass[aps,prd,groupedaddress,graphicx,nofootinbib]{revtex4}
\usepackage{amsmath,amssymb,graphics,graphicx,color,epsf}
\usepackage{subfigure}
\usepackage[scanall]{psfrag}  % the scanall option enables \tex inside EPS files
\usepackage{tikz}

\begin{document}

\title{Topological Eternal Hilltop Inflation and the Swampland Criteria}
%\author{Kazunori Kohri and Chia-Min Lin}

%\author{Kazunori Kohri$^{1}$}\email{kohri@post.kek.jp}
%\author{C. S. Lim$^{2}$}\email{lim@lab.twcu.ac.jp}
\author{Chia-Min Lin}%\email{cmlin@ncut.edu.tw}
%\author{Kin-Wang Ng$^{2,3}$}%\email{cmlin@ncut.edu.tw}
%\author{Kingman Cheung$^{4,5,6}$}%\email{cmlin@ncut.edu.tw}
\affiliation{Fundamental Education Center, National Chin-Yi University of Technology, Taichung 41170, Taiwan}
%\affiliation{$^2$Institute of Physics, Academia Sinica, Taipei 11529, Taiwan}
%\affiliation{$^3$Institute of Astronomy and Astrophysics, Academia Sinica, Taipei 11529, Taiwan}
%\affiliation{$^4$Depaertment of Physics, National Tsing Hua University, Hsinchu 300, Taiwan}
%\affiliation{$^5$Physics Division, National Center for Theoretical Sciences, Hsinchu 300, Taiwan}
%\affiliation{$^6$Division of Quantum Phases and Devices, School of Physics, Konkuk University, Seoul 143-701, Republic of Korea}
%\affiliation{$^1$Cosmophysics Group, Theory Center, IPNS
%KEK, and The Graduate University for Advanced Studies (Sokendai), 1-1
%Oho, Tsukuba 305-0801, Japan}
%\affiliation{$^{2,3}$Department of Physics, Kobe University, Kobe 657-8501, Japan}
%\affiliation{$^2$Department of Mathematics, Tokyo Woman's Christian University, Tokyo 167-8585, Japan}
%\affiliation{$^3$Department of Physics, Chuo University, Bunkyo-ku, Tokyo 112, Japan}

%\baselineskip 14pt

\date{Draft \today}

\begin{abstract}
The swampland criteria put significant constraints on inflation models. In this paper, I show that hilltop inflation on the brane may provide an initial condition for hilltop inflation and produce topological eternal inflation without violating the swampland distance and refined de Sitter criteria. This shows a way that string theory may not be incompatible with eternal inflation.
\end{abstract}
\maketitle
\large
\baselineskip 18pt
\section{Introduction}

Cosmic inflation \cite{Starobinsky:1980te,Brout:1977ix,Guth:1980zm,Sato:1980yn} is regarded as the standard model for the very early universe by many. The simple idea that the universe underwent a period of accelerated expansion solves problems of conventional hot big bang theory. For example, it solves the flatness problem by making the universe much bigger than our observable universe. It solves the horizon problem by shrinking the comoving horizon. It solves the unwanted relics (like the monopoles or other topological defects) problem by diluting them. Inflation generates superhorizon density perturbations from quantum fluctuations in vacuum. It is difficult to explain the observed acoustic peaks in the spectrum of cosmic microwave background without superhorizon perturbations \cite{Dodelson:2003ip}. 

Inflation is usually realized by using a scalar field $\phi$ with its potential $V(\phi)$ and different models correspond to different potential forms. The equation of motion of a homogeneous scalar field in an expanding universe is 
\begin{equation}
\ddot{\phi}+3H\dot{\phi}+V^{\prime}=0.
\label{motion}
\end{equation}
Here the Hubble parameter $H$ provides a friction term $3H\dot{\phi}$ which may make the inflaton rolling slowly. The slow-roll parameters are given by
\begin{equation}
\epsilon \equiv \frac{1}{2}M_P^2 \left( \frac{V^\prime}{V}\right)^2,
\label{epsilon}
\end{equation}
and
\begin{equation}
\eta \equiv M_P^2\frac{V^{\prime\prime}}{V},
\label{eta}
\end{equation}
where $M_P \simeq 2.4 \times 10^{18}$ GeV is the reduced Planck mass. Slow-roll inflation happens when 
\begin{equation}
\epsilon \ll 1,\;\;\;\mbox{and}  \;\;\;  |\eta| \ll 1.
\label{slow}
\end{equation}
The spectral index is given by
\begin{equation}
n_s=1-6\epsilon+2\eta.
\label{index}
\end{equation}
The tensor-to-scalar ratio is 
\begin{equation}
r=16\epsilon.
\label{tensor}
\end{equation}
The number of e-folds is given by
\begin{equation}
N=\frac{1}{M_P^2}\int^\phi_{\phi_e}\frac{V}{V^\prime}d\phi,
\label{efolds}
\end{equation}
where $\phi_e$ denotes field value at the end of inflation. 

Since there is no evidence of primordial non-Gaussianity in recent experiments \cite{Akrami:2018odb}, I will only consider single-field inflation models in the following discussions. The idea of the swampland conjectures (see \cite{Palti:2019pca} for a review) can be used to distinguish different inflation models. This approach has been recently worked in \cite{Kehagias:2018uem,Achucarro:2018vey,Kinney:2018nny,Das:2018hqy,Ashoorioon:2018sqb,Brahma:2018hrd,Lin:2018rnx,Yi:2018dhl,Motaharfar:2018zyb,Lin:2018kjm,Holman:2018inr,Das:2018rpg,Park:2018fuj,Cheong:2018udx,Sabir:2019wel,Sabir:2019bsh,Channuie:2019vsp,Sabir:2019xwk,Wu:2019xtv,Kamali:2019xnt,Heckman:2018mxl,Heckman:2019dsj,Garg:2018zdg}.

String theory unifies all forces of Nature at a quantum level. It admits a huge number of vacua called the landscape after compactification \cite{Bousso:2000xa,Susskind:2003kw,Douglas:2003um,Kachru:2003aw,Denef:2004ze,Douglas:2006es}. However, it is proposed that there is an even much larger varieties of low energy effective field theories (EFTs) located outside the landscape called the swampland. Although theories in the swampland may look self-consistent, it is believed to be contradictory to string theory in particular and quantum gravity in general. The idea is to motivate general properties of quantum gravity. By inductive reasoning, it may be possible to come up with some criteria or conjectures called the swampland criteria. The swampland criteria is then used to differentiate a theory in the landscape from one in the swampland. These criteria are not proved yet, however it is possible build a network of signposts or evidences for them\footnote{It is also possible that the criteria are false \cite{Kallosh:2019axr,Akrami:2018ylq}.}. For a scalar field $\phi$ (which can be applied to the inflaton field), there are \cite{Ooguri:2006in,Obied:2018sgi,Ooguri:2018wrx,Ooguri:2016pdq}\footnote{Recently, a new swampland condition is proposed in \cite{Bedroya:2019snp} based on trans-Planckian censorship (see also \cite{Dvali:2013eja,Dvali:2018jhn}). It will not be addressed in this paper.}.
\begin{itemize}
\item \textbf{The distance conjecture:}
\begin{equation}
\frac{\Delta \phi}{M_P} < \mathcal{O}(1),
\label{eq1}
\end{equation}
\item \textbf{(Refined) de Sitter conjecture:}
\begin{equation}
M_P \frac{|V^\prime|}{V}>c \sim \mathcal{O}(1)\left(\mbox{ or } M_P^2 \frac{V^{\prime\prime}}{V} < -c^\prime \sim -\mathcal{O}(1)\right),
\label{eq2}
\end{equation}
\end{itemize}
The distance conjecture states that scalar field excursions in reduced Planck units in field space are bounded from above \cite{Ooguri:2016pdq}\footnote{The original argument for the distance conjecture in \cite{Ooguri:2006in} based on the infinite tower of states becoming light gives $\Lambda_{QG}=\Lambda_0 e^{-\lambda \Delta \phi}$, where $\Lambda_{QG}$ is the quantum gravity cut-off, $\Lambda_0$ is the original naive cut-off of the EFT, and $\lambda$ is argued to be of order unity in Planck units. It was pointed out in \cite{Scalisi:2018eaz} that even if $\lambda = 1$ the distance conjecture can be relaxed for inflation models to roughly $\Delta \phi \sim 10 M_P$ because the scale of inflation is smaller the the scale of quantum gravity. See also \cite{Andriot:2020lea} and \cite{Gendler:2020dfp} for further discussions of the bounds of $\lambda$.}. The de Sitter conjecture in \cite{Obied:2018sgi} does not contain the part in the parenthesis and it is later refined in \cite{Ooguri:2018wrx} (see also \cite{Garg:2018reu,Denef:2018etk,Andriot:2018wzk,Andriot:2018ept,Roupec:2018mbn,Conlon:2018eyr,Dasgupta:2018rtp,Danielsson:2018qpa}). The refined de Sitter conjecture allows a scalar field with a potentail maximum, namely a hilltop to exist. 

These conjectures violate slow-roll condition Eq.~(\ref{slow}), thus the simplest slow-roll inflation is in the swampland. Therefore we are motivated to consider more complicated scenarios. One possibility is to consider higher dimensional inflationary models where matter fields are confined on a lower dimensional brane while gravity can propagate in the bulk \cite{Maartens:1999hf, Lin:2018kjm}. The motivation to consider this scenario is that in the early universe the Hubble parameter can be enhanced and hence the friction term in Eqs.~(\ref{motion}) is enhanced, which makes the inflaton field to be slow-rolling even without a very flat potential.

If the field value during inflation is smaller than Planck scale (so-called small field inflation), we have $\epsilon \ll |\eta|$, and the spectral index $n_s$ can be approximately given by
\begin{equation}
n_s = 1+2 \eta.
\end{equation}
Since a red spectrum ($n_s < 1$) is observed \cite{Akrami:2018odb}, it means $\eta <0$, namely hilltop inflation. However, the refined de Sitter conjecture suggests $\eta < -1$ which does not fit the allowed range of $n_s \sim 0.96$. This problem can be solved if we consider hilltop inflation on a brane \cite{Lin:2018rnx} as will be presented in Section \ref{section4}.

If hilltop inflation (on a brane) is favored by the swampland criteria and current observation, one may wonder whether there is a fine-tuning problem for a scalar field to sit near the hilltop in order for inflation to occur. The answer to this question is that hilltop inflation is an eternal inflation \cite{Boubekeur:2005zm}. Actually, there are a few different types of eternal inflation mechanisms. The first kind of eternal inflation happens in the original inflation model proposed by Guth \cite{Guth:1980zm,Guth:1982pn}. It is called old inflation nowadays. It is based on a first-order phase transition. The phase transition does not end everywhere in the universe therefore inflation continues eternally. The second one was originally proposed in the framework of stochastic/chaotic inflation \cite{Starobinsky:1986fx,Linde:1986fc,Aryal:1987vn} which is based on the random walk of the inflaton field and I will refer to this mechanism the stochastic eternal inflation. This could also happen in hilltop inflation \cite{Barenboim:2016mmw}. The third type of eternal inflation is called topological eternal inflation \cite{Vilenkin:1994pv,Linde:1994hy,Linde:1994wt} which happens to hilltop inflation \cite{Boubekeur:2005zm}. The question concerning the initial condition of hilltop inflation is actually two-folded. The first question is how did inflation happen? The second question is once inflation happened, whether it is eternal.

The question about whether eternal inflation can be realized under the swampland criteria was investigated in \cite{Matsui:2018bsy,Dimopoulos:2018upl,Kinney:2018kew,Brahma:2019iyy,Wang:2019eym,Rudelius:2019cfh,Blanco-Pillado:2019tdf}. In \cite{Matsui:2018bsy,Dimopoulos:2018upl}, the original de Sitter swampland conjecture were shown to be incompatible with eternal inflation. After the refined swampland de Sitter conjected was taken into account, it is shown in \cite{Kinney:2018kew} that stochastic eternal inflation can happen in hilltop inflation consistent with the swampland criteria. However, an objection is proposed in \cite{Brahma:2019iyy} via a perturbativity constraint of curvature perturbation. In \cite{Wang:2019eym} it is argued that there is a tension between entropy considerations and stochastic eternal inflation when the swampland criteria are considered. It is also proposed that in the case of stochastic eternal inflation there is a graceful exit problem for both large and small field inflation. In \cite{Rudelius:2019cfh}, the Fokker-Planck equation for stochastic eternal inflation is solved either analytically or numerically for some inflation models, but there is no conclusion that stochastic eternal inflation is in the swampland. In \cite{Blanco-Pillado:2019tdf}, a new kind of eternal inflation with an eternally inflating bubble wall is proposed in order to evade the swampland criteria. 

In this paper, I will investigate whether eternal inflation can be realized under the swampland criteria from the view point of topological eternal inflation.

\section{Topological Eternal Inflation}
As mentioned in the Introduction, the unwanted relics problem is solved by inflation via dilution, namely they are inflated away. They still exist somewhere even after inflation ends in our observable universe. If a topological defect (such as monopoles, cosmic strings, domain walls) can inflate, it means the corresponding potential for the scalar field can support slow-rolling. Since the topological defect still exist, the scalar field at positions close to the topological defect has to be near the top of the potential, therefore it will still be inflating and the process continues eternally \cite{Vilenkin:1994pv,Linde:1994hy,Linde:1994wt}. Thus hilltop inflation is eternal without stochastic random walk of the inflaton field value during inflation. The inflaton field is forced to stay near the maximum of the potential at the center of the topological defect for topological reasons.

Let us start our discussion from a simple double-well potential for simplicity 
\begin{equation}
V(\phi)=\frac{1}{4}\kappa (\phi^2-M^2)^2,
\label{qudratic}
\end{equation}
where $\phi$ is presumably an inflaton field.
This potential is of a hilltop form. Near the hilltop when $\phi \sim 0$, the potential can be approximated as 
\begin{equation}
V=V_0-\frac{1}{2}m^2 \phi^2,
\end{equation}
where $V_0 = \frac{1}{4} \kappa M^4$ and $m^2 = \kappa M^2$.
There is a $Z_2$ symmetry under the transformation $\phi \rightarrow -\phi$. After symmetry breaking, domain walls are produced through Kibble mechanism \cite{Kibble:1976sj} if gravitational effects are ignored.  The thickness of the domain wall, $\delta$ is determined by the balance between the gradient and potential energy, $(\frac{M}{\delta})^2 \sim V_0$. This implies
\begin{equation}
\delta \sim \frac{M}{\sqrt{V_0}}.
\end{equation} 
The condition to have the size of the domain wall bigger than the universe, is to have the thickness of the domain wall larger than the Hubble horizon, namely, $\delta > 1/H$. In this case, the Hubble parameter can be obtained from the Friedmann equation $V_0 \sim H^2 M_P^2$. Therefore we have
\begin{equation}
\delta \sim \frac{M}{\sqrt{V_0}} > \frac{M_P}{\sqrt{V_0}}.
\end{equation}
Therefore the condition is basically \cite{Vilenkin:1994pv}
\begin{equation}
M \gtrsim M_P.
\label{condition}
\end{equation}
In this case, it seems that the energy density in the universe is dominated by the potential energy and an equally amount of gradient energy. However this cannot be the full story because in this case, the effects of gravity cannot be ignored. When taking gravitational effects into account, the topological defects inflate and the gradient energy would be diluted \cite{Linde:1994hy}. 

The condition Eq.~(\ref{condition}) can also be obtained from the slow-roll condition Eq.~(\ref{eta}), namely 
\begin{equation}
|\eta| = \left| -\frac{\kappa M^2}{\frac{1}{4}\kappa M^4} \right| = \left| -\frac{4}{M^2} \right| \ll 1. 
\label{condition2}
\end{equation}
Therefore if Eq.~(\ref{condition}) is satisfied, the topological defects are not static and would be inflating. This provides a sufficient condition for hilltop inflation to occur and we can explain why inflation had happened and once it had happened, it would be eternal for topological reasons. Note that as pointed out in \cite{Linde:1994hy}, it is not a necessary condition, because if we consider some chaotic initial condition for the inflaton field, inflation could happen anyway if the field inside some Hubble patch with size $\sim H^{-1}$ happened to be on the hilltop.

However, condition Eq.~(\ref{condition}) violates the swampland distance conjecture and condition Eq.~(\ref{condition2}) violates the refined swampland de Sitter conjecture. Therefore it seems that topological eternal inflation is in the swampland. I will address this issue further in section \ref{section4}.

\section{More hilltop inflation models}

The potential form of a hilltop quartic model is given by
\begin{equation}
V=V_0-\frac{1}{4}\lambda \phi^4.
\end{equation}
In Fig.8 of the Planck 2018 results \cite{Akrami:2018odb}, the predictions of hilltop quartic model is compared with experimental constraints. As can be noticed in the figure, the predicted spectral index $n_s$ is not in the range allowed by experimental constraints unless the tensor-to-scalar ratio becomes larger. This implies the required inflaton field value approaches Planck scale, namely the model is leaving the regime of small field inflation. This can be seen from Eq.~(\ref{epsilon}) and Eq.~(\ref{efolds}) and is known as Lyth bound \cite{Lyth:1996im}. However, in this parameter regime, the field value when the observable universe leaves horizon will not be sitting near the hilltop and moreover there is a graceful exit problem to end inflation \cite{Kallosh:2019jnl}. I provide an analysis of hilltop quartic model in the Appendix. Actually hilltop quartic model violates the swampland de Sitter conjecture Eq.~(\ref{eq2}) at the top of the potential.

Nevertheless, there are more hilltop inflation models \cite{Kohri:2007gq} than the hilltop quartic model (and hilltop quadratic model) and they can fit the experimental constraints without leaving the regime of small field inflation. In \cite{Kohri:2007gq}, an effective inflaton potential during inflation of the following form is considered with a positive $\lambda$
\begin{eqnarray}
V(\phi) &=& V_0 \pm \frac{1}{2}m^2 \phi^2 - \lambda \frac{\phi^p}{M_P^{p-4}}+\cdots   \\
            &\equiv& V_0 \left( 1+\frac{1}{2}\eta_0 \frac{\phi^2}{M_P^2} \right) - \lambda \frac{\phi^p}{M_P^{p-4}}+\cdots,
\label{hilltop}
\end{eqnarray} 
with 
\begin{equation}
\eta_0=\frac{\pm m^2 M_P^2}{V_0}.
\end{equation}
The dots $\cdots$ in the potential represents higher order terms which are ignored during inflation because a small field inflation model with $\phi \ll M_P$ is considered. However, they are important to stabilize the potential after inflation ends. Depending on whether $\eta_0 \leq 0$ or $\eta_0 > 0$ and $p>2$ or $p<0$, three types of hilltop inflation models can be considered. Note that $p$ is not necessarily an integer. By taking the limit $p \rightarrow 0$ with $\lambda p$ fixed, the model can describe F- and D-term inflation \cite{Copeland:1994vg,Stewart:1994ts,Dvali:1994ms,Binetruy:1996xj,Halyo:1996pp} and their hilltop modification \cite{Lin:2006xta, Lin:2008ys, Lin:2009yt}. The potential Eq.~(\ref{hilltop}) may look complicated, however it can still be analyzed. For example, the field value during inflation is given by \cite{Kohri:2007gq}
\begin{equation}
\left( \frac{\phi}{M_P} \right)^{p-2}=\left( \frac{V_0}{M_P^4}\right)\frac{\eta_0 e^{(p-2)\eta_0N}}{\eta_0 x + p\lambda(e^{(p-2)\eta_0 N}-1)},
\end{equation}
where
\begin{equation}
x \equiv \left(\frac{V_0}{M_P^4} \right) \left( \frac{M_P}{\phi_e}\right)^{p-2}.
\end{equation}
The spectral index is 
\begin{equation}
n_s=1+2\eta_0 \left[ 1-\frac{\lambda p (p-1) e^{(p-2)\eta_0 N}}{\eta_0 x + p\lambda(e^{(p-2)\eta_0 N}-1)} \right].
\end{equation}

If the potential has a hilltop, in causally disconnected regions of the universe the scalar field value could be on either side of the hilltop. If gravitational effects or inflation is ignored, topological defects still forms even if the symmetry is not exact. Since the potential given in Eq.~(\ref{hilltop}) is more complicated than the potential in Eq.~(\ref{qudratic}), it is not so easy to determine the vacuum expectation value (vev) of $\langle \phi \rangle \equiv M$. Actually for hybrid inflation, usually the vev of $\phi$ after inflation is zero. In this case, I am referring to the vev on the other side of the hilltop potential. It is possible that $\phi=M$ corresponds to a false vacuum with non-vanishing potential, but it does not change our conclusion. In any case, we can estimate $M$ as the following. Since the terms corresponding to the dots $\cdots$ in Eq.~(\ref{hilltop}) are suppressed in $M_P$, we can expect those terms to become important when $\phi \rightarrow M_P$ therefore $M \gtrsim M_P$ as in Eq.~(\ref{condition}). We can also use dimensional analysis to the slow-roll parameter Eq.~(\ref{eta}) to obtain
\begin{equation}
|\eta| \sim \frac{M_P^2}{M^2} < 1,
\end{equation}
which again suggests $M>M_P$. Although these are rough estimation, it is likely that the swampland distance conjecture given by Eq.~(\ref{eq1}) is still violated.

\section{Hilltop inflation on a brane}
\label{section4}
There is a way out from the condition given by Eq.~(\ref{condition}). The derivation of the condition is based on the conventional Friedmann equation $\rho = 3 H^2 M_P^2$. However, it can be modified  in the early universe if we consider a braneworld scenario where our four-dimensional world is a 3-brane embedded in a higher-dimensional bulk. By assuming $\rho \sim V$ the Friedmann equation can be modified as \cite{Cline:1999ts,Csaki:1999jh,Binetruy:1999ut,Binetruy:1999hy,Freese:2002sq,Freese:2002gv,Maartens:1999hf}
\begin{equation}
H^2=\frac{V}{3M_P^2}\left( 1+\frac{V}{2\Lambda}\right),
\label{brane}
\end{equation}
where 
\begin{equation}
\Lambda \equiv 6\pi \frac{M_5^6}{M_P^2},
\end{equation}
and $M_5$ is the reduced Planck scale  in five dimensions. The nucleosynthesis limit implies that $\Lambda \gtrsim (1 \mbox{ MeV})^4 \sim (10^{-21})^4$ \cite{Cline:1999ts}. A more stringent constraint, $M_5 \gtrsim 10^5$ TeV, can be obtained by requiring the theory to reduce to Newtonian gravity on scales larger than 1 mm, this corresponds to $\Lambda \gtrsim 5.0 \times 10^{-53}$ \cite{Randall:1999ee, Randall:1999vf}. Since the lower bound of $\Lambda$ is quite small, it is easy to obtain $V_0 /\Lambda \gg 1$ when the potential energy density of the universe is given by $V \sim V_0$ near the hilltop, so that the brane effect is significant. In this case, we have $H^2=V_0^2 /6M_P^2 \Lambda$. Therefore the Hubble horizon is given by
\begin{equation}
\frac{1}{H}=\frac{M_P \sqrt{6\Lambda}}{V_0}.
\end{equation}
The condition $\delta > 1/H$ becomes
\begin{equation}
M > M_P \sqrt{\frac{6\Lambda}{V_0}}.
\end{equation}
Since we assume $V_0 /\Lambda \gg 1$, the swampland distance conjecture is satisfied. In this case the slow-roll parameters are modified into \cite{Lin:2018rnx} 
\begin{equation}
\epsilon = \frac{M_P^2}{2}\left(\frac{V^\prime}{V_0}\right)^2\frac{1}{\left(\frac{V_0}{4\Lambda}\right)},
\end{equation}
and 
\begin{equation}
\eta=M_P^2 \left(\frac{V^{\prime\prime}}{V_0} \right)\frac{1}{\left(\frac{V_0}{2\Lambda}\right)}.
\end{equation}
We can see that due to the suppression $V_0 /\Lambda \gg 1$, the swampland (refined) de Sitter conjecture Eq.~(\ref{eq2}) can be satisfied. The physical reason behind this results can be seen from Eqs.~(\ref{motion}) and (\ref{brane}). In the early universe when $V_0 /\Lambda \gg 1$ during inflation, the Hubble parameter is enhanced and hence the friction term in the equation of motion of the inflaton field is enhanced, which makes the inflaton field to be slow-rolling even without a very flat potential.

\section{Conclusion and Discussion}
\label{con}

In this paper I have shown that hilltop inflation on the brane can reduce the required vacuum expectation value of the inflaton field in order to produce a topological defect as large as a universe. This in turn sets the initial condition of inflation and produce topological eternal inflation without contradict to the swampland distance and refined de Sitter criteria.

\appendix
\section{analysis of hilltop quartic model}

In this appendix, I provide an analysis to obtain relevant results of hilltop quartic model by hand for those who may be interested. It goes without saying that the calculation can also be done by numerical methods, but I believe it is more inspiring to do it by hand.
In the following, I always set the reduced Planck mass $M_P=1$. For notational simplicity throughout the calculation, I will rewrite the potential as
\begin{equation}
V=V_0-\frac{1}{4}\lambda \phi^4 \equiv V_0-\frac{V_0}{\mu^4}\phi^4.
\end{equation}
Therefore we have
\begin{equation}
V^\prime=-4\frac{V_0}{\mu^4}\phi^3, \;\;\; V^{\prime\prime}=-12\frac{V_0}{\mu^4}\phi^2.
\end{equation}
From Eq.~(\ref{epsilon}), we have
\begin{equation}
\epsilon = 8 \frac{\phi^6}{\mu^8}\frac{1}{\left( 1-\frac{\phi^4}{\mu^4} \right)^2}.
\end{equation}
From Eq.~(\ref{eta}), 
\begin{equation}
\eta=-12\frac{\phi^2}{\mu^4}\frac{1}{\left( 1-\frac{\phi^4}{\mu^4} \right)}.
\end{equation}
Inflation ends at $\phi_e$ when $\eta=-1$, hence
\begin{equation}
\phi_e^4-12\phi_e^2-\mu^4=0.
\end{equation}
Let $X_e \equiv \phi_e^2$, we have
\begin{equation}
X_e^2-12X_e-\mu^4=0.
\end{equation}
Therefore 
\begin{equation}
X_e=6+\sqrt{36+\mu^4},
\end{equation}
where the minus solution is ignored.

\begin{figure}[t]
\centering
\includegraphics[width=0.7\columnwidth]{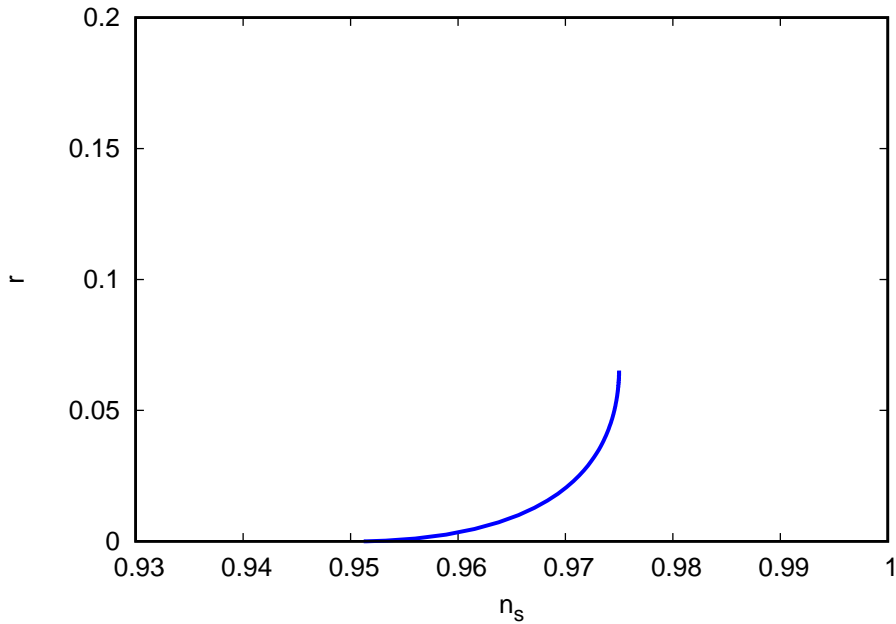}
 \caption{The spectral index $n_s$ versus the tensor-to-scalar ratio $r$ for hilltop quartic model.}
\label{fig1}
\end{figure}

From Eq.~(\ref{efolds}), the number of e-folds when our observable universe leaves horizon is given by 
\begin{eqnarray}
N &=& \int_{\phi_e}^\phi \frac{V_0-\frac{V_0}{\mu^4}\phi^4}{-4\frac{V_0}{\mu^4}\phi^3}   \\
    &=& \frac{1}{8}\phi^2+\frac{1}{8}\frac{V_0}{\phi^2}-\frac{1}{8}X_e-\frac{1}{8}\frac{\mu^4}{X_e}   \\
    &=&  60,
\end{eqnarray}
where $N=50$ can also be used. Therefore we have
\begin{equation}
\phi^4-\left( X_e+\frac{\mu^4}{X_e}+480 \right)\phi^2+\mu^4=0.
\end{equation}
Let $X \equiv \phi^2$, we obtain
\begin{equation}
X=\frac{\left( X_e+\frac{\mu^4}{X_e}+480 \right)-\sqrt{\left( X_e+\frac{\mu^4}{X_e}+480 \right)^2-4\mu^4}}{2}.
\end{equation}
We have chosen the solution which corresponds to $X<X_e$.  Note that $X(\mu)$ and $X_e(\mu)$ are functions of $\mu$. In terms of $X$, we can write the spectral index as 
\begin{equation}
n_s=1+2\eta-6\epsilon=1-24\frac{X}{\mu^4}\frac{1}{\left(1-\frac{X^2}{\mu^4} \right)}-48\frac{X^3}{\mu^8}\frac{1}{\left(1-\frac{X^2}{\mu^4} \right)^2}.
\end{equation}
The tensor-to-scalar ratio is obtaiend from Eq.~(\ref{tensor}) as
\begin{equation}
r=128\frac{X^3}{\mu^8}\frac{1}{\left( 1-\frac{X^2}{\mu^4} \right)^2}.
\end{equation}
We can therefore plot $r$ versus $n_s$, by using $\mu$ as a parameter. It is given in Fig.~\ref{fig1}. I bypass the imposing of cosmic microwave background (CMB) normalization, which would give a relation between $V_0$ and $\mu$.

%\section*{Acknowledgement}
%This work is supported by the Ministry of Science and Technology (MOST) of Taiwan under grant number MOST 106-2112-M-167-001. 

\end{document}